\documentclass[12pt]{article}
\usepackage{graphics,epsfig}
\begin{document}


\begin{center}
\vspace*{0.5cm}
{\large{\bf Chiral Dynamics in Nuclear Systems}\footnote
{Presented at CHIRAL 02, Kyoto, Japan}\\}
\vspace{0.5cm}
{Wolfram WEISE\footnote{Supported in part by BMBF, GSI and JSPS}\\} 
\vspace{0.5cm}
{$ECT^*$, I-38050 Villazzano (Trento), Italy \\
and \\
Physics Department, Technical University of Munich, D-85747 Garching, Germany}
\end{center}

\abstract{
A survey is given on selected topics concerning the role of spontaneous
chiral symmetry breaking in low-energy QCD, and its 
dynamical implications for nuclear systems. This includes aspects of chiral
thermodynamics (the temperature and density dependence of the chiral 
condensate). It also includes an update on the theory of low-energy
(s-wave) pion-nuclear interactions relevant for deeply-bound states of
pionic atoms and the quest for possible fingerprints of chiral symmetry 
restoration in nuclear systems.}

\section{Introduction}
This presentation deals with the low-energy dynamics governed by the 
approximate chiral symmetry of QCD.
Conceptually and by analogy, low-energy QCD has many features in common with
condensed matter physics. The QCD ground state (or vacuum) has a complex
structure. It hosts strong condensates of quark-antiquark pairs and gluons. The
quark condensate $\langle \bar{q}q \rangle$, i.~e. the ground state expectation
value of the scalar quark density, plays a particularly important role. 
It represents an order parameter of
spontaneously broken chiral symmetry. Its behaviour as a function of
temperature and baryon density is therefore of prime interest. The possible
restoration of chiral symmetry above some critical temperature and density (the
chiral phase transition) is in fact one of the key issues that motivates 
the field of high-energy heavy-ion collisions.

We begin by briefly summarizing basics of QCD and chiral symmetry.
While the focus in the present paper is primarily on the confinement 
phase of QCD
which includes ordinary hadrons and nuclei, the global QCD phase structure, its
relevant scales and symmetry breaking patterns, should always be kept in mind
as we go along. The thermodynamics of the chiral condensate will be an 
important part of our discussion. The last section highlights the recent 
revival of interest in low-energy pion-nucleus interactions and the quest for 
fingerprints of "chiral restoration", inspired by new accurate 
experimental data on
deeply bound pionic atoms with heavy nuclei. Another subject of great interest
is the in-medium evolution of the scalar-isoscalar $(J^{\pi} = 0^+, I=0)$
two-pion spectral function \cite{HK}. 
This topic will not be covered in detail here as it is
well represented in other parts of these proceedings. 

\section{QCD and chiral symmetry}
\subsection{Basics}
Our framework is QCD in the sector of the lightest ($u$-, $d$-) quarks. 
They form a flavour $N_f = 2$ (isospin)
doublet with "bare" quark masses of less than 10 MeV.
The flavour (and colour) components of the quarks are collected in 
the Dirac fields $\psi (x) = (u(x), d(x))^T$. The QCD Lagrangian is
\begin{equation}
{\cal L}_{QCD} = \bar{\psi} (i \gamma_{\mu} D^{\mu} - m) \psi - \frac{1}{2} Tr
(G_{\mu \nu} G^{\mu \nu}),
\end{equation}
with the $SU(3)_{color}$ gauge covariant derivative $D^{\mu}$ and the gluonic
field tensor $G^{\mu \nu} = (i/g) [ D^{\mu}, D^{\nu} ]$.
The $2 \times 2$
matrix $m = diag (m_u, \, m_d)$ contains the light quark masses. 
The strange quark is more than
an order of magnitude heavier ($m_s \sim$ 150 MeV), but still sometimes 
considered
"light" on the typical GeV scales of strong interaction physics. 
The heavy quarks ($Q = c$, $b$ and $t$) can be ignored in the present
context. Their masses are separated from those of the light quarks by several
orders of magnitude. 

At short distance scales, $r < 0.1$ fm, corresponding to momentum transfers
above several GeV/c, QCD is a perturbative theory of pointlike quarks and
gluons. The rules for their dynamics are set by local gauge invariance under
$SU(3)_{colour}$. At large distance scales, $r> 1$ fm, corresponding to low
energies and momenta relevant to most of nuclear physics, QCD is realised as a
theory of pions and nucleons (and possibly other heavy, almost static 
hadrons). Their low-energy dynamics is governed by the spontaneous breaking 
of an approximate symmetry of QCD: {\it Chiral Symmetry}.

Consider QCD in the limit of massless quarks, setting $m=0$ in eq.(1). 
In this limit, the QCD Lagrangian has a global symmetry related to 
the conserved right-
or left-handedness (chirality) of zero mass spin $1/2$ particles. 
Introducing right- and left-handed quark fields,
\begin{equation}
\psi_{R,L} = \frac{1}{2} (1 \pm \gamma_5) \psi ,
\end{equation}
we observe that separate global unitary transformations
\begin{equation}
\psi_R \to \exp [i \theta^a_R \frac{\tau_a}{2}] \, \psi_R, \hspace{1,5cm} \psi_L
\to \exp [i \theta^a_L \frac{\tau_a}{2}] \, \psi_L ,
\end{equation}
with $\tau_a (a= 1,2,3)$ the generators of (isospin) $SU(2)$, 
leave ${\cal L}_{QCD}$
invariant in the limit $m \to 0$. This is the chiral $SU(2)_R \times
SU(2)_L$ symmetry of QCD. It implies six conserved Noether currents,
$J^{\mu}_{R, a} = \bar{\psi}_R \gamma^{\mu} \frac{\tau_a}{2} \psi_R$ and
$J^{\mu}_{L, a} = \bar{\psi}_L \gamma^{\mu} \frac{\tau_a}{2} \psi_L$, with
$\partial_{\mu} J^{\mu}_R = \partial_{\mu} J^{\mu}_L = 0 $. It is common to
introduce the vector current
\begin{equation}
V^{\mu}_a = J^{\mu}_{R,a} + J^{\mu}_{L,a} = \bar{\psi} \gamma^{\mu}
\frac{\tau_a}{2} \psi ,
\end{equation}
and the axial current,
\begin{equation}
A^{\mu}_a (x) = J_{R,a} - J_{L,a} = \bar{\psi} \gamma^{\mu} \gamma_5 \frac{\tau_a}{2} \psi .
\end{equation}
Their corresponding charges,
\begin{equation}
Q^V_a = \int d^3 x 
~\psi^{\dagger} (x) \frac{\tau_a}{2} \psi (x), \hspace{1,5cm}
Q^A_a = \int d^3 x 
~\psi^{\dagger} (x) \gamma_5 \frac{\tau_a}{2} \psi (x) ,
\end{equation}
are, likewise, generators of $SU(2) \times SU(2)$.

\subsection{Spontaneous chiral symmetry breaking}
There is evidence from hadron spectroscopy that the chiral  
$SU(2) \times SU(2)$ symmetry 
of the QCD Lagrangian (1) with $m = 0$ is
spontaneously broken: for dynamical reasons of non-perturbative
origin, the ground state (vacuum) of QCD has lost part of the symmetry of the
Lagrangian. It is symmetric only under the subgroup $SU(2)_V$ generated by 
the vector charges $Q^V$. This is the well-known isospin symmetry seen in 
spectroscopy and dynamics.

If the ground state of QCD were symmetric under chiral $SU(2) \times SU(2)$,
both vector and axial charge operators would annihilate the vacuum: $Q^V_a |0 \rangle
= Q^A_a |0 \rangle = 0$. This is the Wigner-Weyl realisation of chiral symmetry with
a "trivial" vacuum. It would imply the systematic appearance of parity doublets
in the hadron spectrum. For example, correlation functions of vector and axial
vector currents should be identical, i.~e. $\langle 0 | V^{\mu} V^{\nu} | 0
\rangle = \langle 0 | A^{\mu} A^{\nu} | 0 \rangle$. Consequently, the spectra
of vector $(J^{\pi} = 1^-)$ and axial vector $(J^{\pi} = 1^+)$ mesonic
excitations should also be identical. This degeneracy is not seen in nature:
the $\rho$ meson mass ($m_{\rho} \simeq 0.77$ GeV) is well separated from
that of the $a_1$ meson ($m_{a_1} \simeq 1.23$ GeV). Likewise, the light
pseudoscalar $(J^{\pi} = 0^-)$ mesons have masses much lower than the lightest
scalar $(J^{\pi} = 0^+)$ mesons.

One must conclude $Q^A_a |0 \rangle \neq 0$, that is, chiral symmetry is
spontaneously broken down to isospin: $SU(2)_R \times SU(2)_L \to
SU(2)_V$. This is the Nambu-Goldstone realisation of chiral symmetry.
A spontaneously broken global symmetry implies the existence of a (massless)
Goldstone boson. If $Q^A_a | 0 \rangle \neq 0$, there must be a physical state
generated by the axial charge, $|\phi_a \rangle = Q^A_a | 0 \rangle$, which is
energetically degenerate with the vacuum. Let $H_0$ be the QCD Hamiltonian
(with massless quarks) which commutes with the axial charge. Setting the ground
state energy equal to zero for convenience, we have 
$H_0 |\phi_a \rangle = Q^A_a
H_0 | 0 \rangle = 0$. Evidently $| \phi_a \rangle $ represents three massless
pseudoscalar bosons (for $N_f = 2$). They are identified with the pions.

\subsection{The chiral condensate}
Spontaneous chiral symmetry breaking goes together with a qualitative
re-arrangement of the vacuum, an entirely non-perturbative phenomenon. The
ground state is now populated by scalar quark-antiquark pairs. The
corresponding ground state expectation value $\langle 0 | \bar{\psi} \psi | 0
\rangle$ is called the chiral (or quark) condensate. 
We frequently use the notation
\begin{equation}
\langle \bar{\psi} \psi \rangle = \langle \bar{u} u \rangle + \langle \bar{d} d
\rangle ~~.
\end{equation}
The precise definition of the chiral condensate is:
\begin{equation}
\langle \bar{\psi} \psi \rangle = -i Tr \lim_{y \to x^+} S_F (x,y)
\end{equation}
with the full quark propagator, $S_F (x,y) = -i \langle 0 | {\cal T} \psi (x)
\bar{\psi} (y) | 0 \rangle$ where ${\cal T}$ denotes the time-ordered
product. We recall Wick's theorem which states that ${\cal T} \psi (x)
\bar{\psi} (y)$ reduces to the normal product $ :\psi (x) \bar{\psi} (y):$ plus
the contraction of the two field operators. When considering the perturbative
quark propagator, $S^{(0)}_F (x,y)$, the time-ordered product is taken with
respect to a trivial vacuum for which the expectation value of $ : \bar{\psi}
\psi :$ vanishes. Long-range, non-perturbative physics is then at the origin of
a non-vanishing $\langle : \bar{\psi} \psi  : \rangle$.

(In order to establish the connection between spontaneous chiral symmetry 
breaking and the non-vanishing chiral condensate in a more formal way, 
introduce the pseudoscalar operator
$P_a (x) = \bar{\psi} (x) \gamma_5 \tau_a \psi (x)$ and
derive the (equal-time) commutator relation
$[ Q^A_a, P_b] = - \delta_{ab} \bar{\psi} \psi$
which involves the axial charge $Q^A_a$ of eq.~(6). Taking the ground state
expectation value, we see that $Q^A_a | 0 \rangle \neq
0$ is indeed consistent with $\langle \bar{\psi} \psi \rangle \neq 0$.)

Let $| \pi_a (p) \rangle $ be the state vectors of the Goldstone bosons 
associated
with the spontaneous breakdown of chiral symmetry. Their four-momenta are
denoted $p^{\mu} = (E_p, \vec{p}~)$, and we choose the standard normalization $
\langle \pi_a (p) | \pi_b (p') \rangle = 2 E_p \delta_{ab} (2 \pi)^3 \delta^3
(\vec{p} - \vec{p} \, ')$. Goldstone's theorem
also implies non-vanishing matrix elements of the axial current (5) which
connect $| \pi_a (p) \rangle$ with the vacuum:
\begin{equation}
\langle 0 | A^{\mu}_a (x) | \pi_b (p) \rangle = i p^{\mu} f \delta_{ab} e^{-ip
\cdot x} ,
\end{equation}
where $f$ is the pion decay constant (taken here in the chiral
limit, i.~e. for vanishing quark mass). Its physical value
\begin{equation}
f_\pi = (92.4 \pm 0.3) \, MeV
\end{equation}
differs from $f$ by a small correction linear in the quark mass $m_q$.

Non-zero quark masses $m_{u,d}$ shift the mass of the Goldstone boson from zero
to the observed value of the physical pion mass, $m_{\pi}$. The connection between $m_{\pi}$ and the $u-$ and $d-$ quark masses is provided by 
PCAC and the Gell-Mann, Oakes, Renner (GOR) relation
\cite{1}:
\begin{equation}
m^2_{\pi} = - \frac{1}{f^2} (m_u + m_d) \langle \bar{q} q \rangle + 
{\cal O}(m^2_{u,d}) .
\end{equation}
We have set $\langle \bar{q} q \rangle \equiv \langle \bar{u} u \rangle \simeq
\langle \bar{d} d \rangle$ making use of isospin symmetry which is valid to a
good approximation. Neglecting terms of order $m^2_{u,d}$,
identifying $f = f_{\pi} = 92.4$ MeV to this order and inserting $m_u + m_d
\simeq 12$ MeV \cite{2,3} (at a renormalisation scale of order 1 GeV), 
one obtains
\begin{equation}
\langle \bar{q} q \rangle \simeq - (240 \, MeV)^3 \simeq -1.8 \, fm^{-3} .
\end{equation}
This condensate (or correspondingly, the pion decay constant $f_{\pi}$) is a
measure of spontaneous chiral symmetry breaking. The non-zero pion mass, on the
other hand, reflects the explicit symmetry breaking by the small quark masses,
with $m^2_{\pi} \sim m_q$. It is important to note that $m_q$ and $\langle
\bar{q} q \rangle$ are both scale dependent quantities. Only their product
$m_q \langle \bar{q} q \rangle$ is invariant under the renormalisation
group.

The GOR relation in its form (11) with a large order parameter $\langle \bar{q}
q \rangle $ has been challenged in ref.\cite{4} where it was pointed out that
an alternative symmetry breaking scheme may exist in which $\langle \bar{q} q
\rangle$ can be small or even vanishing. 
It would lead to a linear relationship between
$m_{\pi}$ and $m_q$, rather than the quadratic connection $m^2_{\pi} \sim m_q$
characteristic of PCAC. However, at least within two-flavour QCD, 
this issue can be sorted out experimentally by
a detailed analysis of $s$-wave $\pi \pi$ scattering
lengths \cite{5}. Such an analysis confirms quite convincingly that the
"standard" spontaneous symmetry breaking scenario with a large chiral
condensate is indeed very likely the one realised in nature. Possible 
corrections to this picture, when generalised to flavour SU(3), are
still under discussion.

\subsection{Phases of QCD}

The QCD Lagrangian (1) generates a remarkably rich thermodynamics and 
phase structure. Our present knowledge is primarily based on results
from lattice QCD which we briefly summarise here.

Basic order parameters to map out the QCD phase structure are the
Polyakov-Wilson loop and the chiral condensate already mentioned. The Polyakov
loop $L (T)$ is a measure of deconfinement in the limit of static, infinitely
heavy colour sources (quarks):
\begin{equation}
L (T) \sim \exp [-{\cal V} (r \to \infty) / T ] ,
\end{equation}
where ${\cal V}(r)$ is the potential between a static quark and antiquark at distance
$r$. In the confined phase ${\cal V} (\infty) \to \infty$, which implies $L \to
0$. Colour screening at high temperature makes ${\cal V}(r)$ finite at large $r$ so
that $L$ becomes non-zero, indicating deconfinement.

Given the QCD partition function ${\cal Z}$ (see eq.(18)) and the pressure 
$P = T(\partial ln {\cal Z}/\partial V)$, the chiral condensate 
$\langle \bar{\psi} \psi \rangle_T$ at finite temperature
is obtained by taking the derivative of the pressure with respect to
the quark mass:
\begin{equation}
\langle \bar{\psi} \psi \rangle_T \sim \frac{\partial P(T,V)}{\partial m_q} .
\end{equation}
One expects its magnitude to decrease from its $T=0$ value (12) to zero beyond
a critical temperature $T_c$. At $T > T_c$ chiral symmetry is restored in its
Wigner-Weyl realisation. At $T < T_c$ the symmetry is spontaneously broken 
in the low-temperature (Nambu-Goldstone) phase. 
A further interesting and sensitive quantity is the chiral susceptibility
\begin{equation}
{\cal X}_m = \left| \frac{\partial \langle \bar{\psi} \psi \rangle}{\partial
m_q} \right| ,
\end{equation}
corresponding to the second derivative of the free energy with respect to the
quark mass. Whereas the (scale dependent) quantity
$\langle \bar{\psi} \psi \rangle$ is not directly observable as such, the
fluctuations ${\cal X}_m = \langle (\bar{\psi} \psi)^2 \rangle
-  \langle \bar{\psi} \psi \rangle^2$ are, in principle, observable through
their connection with the spectral function of scalar-isoscalar excitations
of the QCD vacuum \cite{HK}.
 
The present qualitative knowledge about the phases of QCD at zero chemical
potential $\mu$ can be summarised as follows. The limiting case of a pure $SU(3)_c$
gauge theory ($m_q \to \infty$ for all quark flavours) exhibits a first-order
deconfinement transition at a critical temperature $T_c \simeq 270$ MeV
\cite{6}. As one moves down to small quark masses and extrapolates to the chiral
limit, $m_q \to 0$, one finds a chiral 1st order transition with critical
temperatures depending on the number of flavours \cite{7}:
\begin{equation}
T_c = (173 \pm 8) \quad\mbox{MeV \hspace{2cm} for}\quad N_f = 2 ,
\end{equation}
\begin{equation}
T_c = (154 \pm 8) \quad\mbox{MeV \hspace{2cm} for}\quad N_f = 3 .
\end{equation}
The critical temperature decreases when increasing the number of
active (light) degrees of freedom.

The physically relevant situation, with chiral symmetry 
explicitly broken by finite
quark masses $m_{u,d} \sim 10$ MeV and $m_s \sim 150$ MeV, is expected to be
between those limiting cases. Along the line $m_u = m_d = 0$ but with $m_s$
steadily increasing, one continues from a domain with a 1st order
transition to a region characterized by a second-order transition restoring chiral
$SU(2) \times SU(2)$ symmetry. For realistic quark masses the transition might
end up as a smooth crossover, signalled by rapid but continuous changes
of quantities such as the chiral condensate. This is shown in Fig.1.

While there is no fundamental reason why the chiral symmetry
restoration and deconfinement transitions should take 
place at the same critical
temperature, lattice results demonstrate that they appear indeed to be
correlated. The chiral condensate $\langle \bar{\psi}
\psi \rangle$ and the susceptibility ${\cal X}_m$ undergo rapid changes at the
same temperature scale as the Polyakov loop $\langle L \rangle$ and its
corresponding susceptibility, ${\cal X}_L \sim \langle L^2 \rangle - \langle L
\rangle^2$.
\begin{figure}[!htb]
  \begin{center}
    \includegraphics*[width=0.5\textwidth, height=5.5cm]{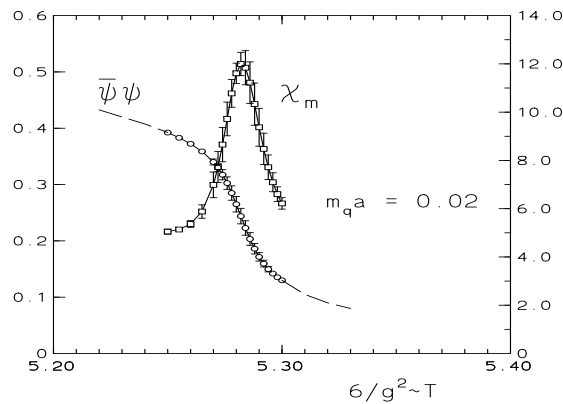}
   \caption{The chiral condensate $\langle
\bar{\psi} \psi \rangle$ and the corresponding susceptibility in 
2-flavour lattice QCD as function of $\beta = 6/g^2
\sim T$, for a quark mass which scales with temperature as $m_q = 0.08 \,
T$. (From ref. \cite{8}.)}\label{fig_1}
  \end{center}
\end{figure}

Non-zero baryon density poses a principal problem in lattice QCD: the Fermion
determinant that appears in the Euclidean path integral representing the 
QCD partition function becomes complex at finite
chemical potential $\mu$ and this prohibits standard numerical algorithms. 
Several strategies are used to overcome this problem, by Taylor expansion
in small $\mu / T_c$ \cite{9}, or by the ``reweighting'' approach 
pursued in ref.\cite{10}. The present results, summarised in Fig.2, 
give a rough picture 
of what to expect for the QCD phase diagram, although with increasing
uncertainties as the chemical potential $\mu$ grows.

\begin{figure}[!htb]
  \begin{center}
    \includegraphics*[width=0.5\textwidth, height=5.5cm]{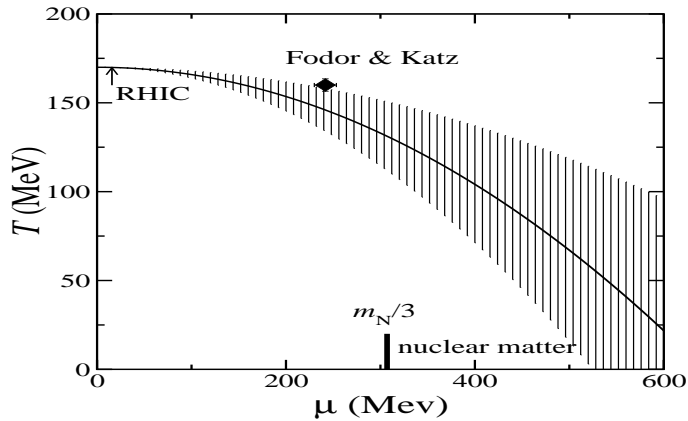}
   \caption{Sketch of the phase diagram as obtained in ref.~\cite{9}. The dot
marks the end point of the first order transition as found in ref.~\cite{10}.}
\label{fig_2}
  \end{center}
\end{figure}

\section{Chiral thermodynamics}
\subsection{Low-energy QCD}

In the hadronic phase of QCD, the thermodynamically active degrees of freedom
are not elementary quarks and gluons but mesons and baryons. Consider as a
starting point the partition function
\begin{equation}
{\cal Z} = Tr \exp\left[-\frac{1}{T}\int_V d^3x~({\cal H}-\mu\rho)\right]~~,
\end{equation}
where ${\cal H}$ is the Hamiltonian density, $\mu$ denotes the chemical 
potential, and $\rho$ the baryon density. The
partition function at $\mu = 0$, expressed in terms of the QCD Hamiltonian
$H$, is
\begin{equation}
{\cal Z} = Tr \exp (-H/T) = \sum_n \langle n | e^{-E_n /T} | n \rangle
\end{equation}
with $(H - E_n) | n \rangle = 0$. Confinement implies that the eigenstates $| n
\rangle$ of $H$ are (colour-singlet) hadrons at $T < T_c$. The low-temperature
physics is then determined by the states of lowest mass in the spectrum $\{ E_n
\}$.

The observed spectrum of low-mass hadrons has a characteristic gap, $\Delta
\sim M_{nucleon} \sim m_{\rho - meson} \sim 1$ GeV, which separates the masses
of all baryons and almost all mesons from the ground state $| 0 \rangle$. On
the other hand, the lightest pseudoscalar mesons are positioned well 
within this gap, for
good reason: as Goldstone bosons of spontaneously broken chiral symmetry they
would start out massless. Explicit symmetry breaking by the masses $m_q$ of the
light quarks introduces perturbations on a scale small compared to
$\Delta$.

The appearance of the gap $\Delta$ is presumably linked to the presence
of the chiral condensate $\langle \bar{\psi} \psi \rangle$ in the QCD ground
state. For example, Ioffe's formula \cite{11}, based on QCD sum rules, connects
the nucleon mass $M_N$ directly with  $\langle \bar{\psi} \psi \rangle$  in
leading order. While this formula is not very accurate and needs to be improved
by including higher order condensates, it nevertheless demonstrates that
spontaneous chiral symmetry breaking plays an essential role in giving the
nucleon its mass.

The condensate  $\langle \bar{\psi} \psi \rangle$ is encoded in the pion decay
constant $f_{\pi}$ through the GOR relation (12). In the chiral limit $(m_q \to
0)$, this $f_{\pi}$ is the only quantity which can serve to define a mass scale
("transmuted" from the QCD scale $\Lambda_{QCD}\sim 0.2$ GeV  
through non-perturbative dynamics). It is common to introduce 
$ 4 \pi f_{\pi} \sim 1$ GeV as the scale characteristic 
of spontaneous chiral symmetry
breaking. This scale is then roughly identified with the spectral
gap $\Delta$. 

As another typical example of how the chiral gap translates into hadron masses,
consider the $\rho$ and $a_1$ mesons. Finite-energy sum rules for
vector and axial vector current-current correlation
functions, when combined with the algebra of these currents as implied by the
chiral $SU(2) \times SU(2)$ group, do in fact connect the $\rho$ and $a_1$
masses directly with the chiral gap \cite{12,13}:
\begin{equation}
m_{a_1} = \sqrt{2} m_{\rho} = 4 \pi f_{\pi} ,
\end{equation}
at least in leading order (that is, in the large $N_c$ limit, and ignoring
decay widths as well as perturbative QCD corrections).

Such mass relations, while not accurate at a quantitative level, give
important hints. Systems characterized by an energy gap usually exhibit
qualitative changes when exposed to variations of thermodynamic 
conditions. For the physics in the hadronic phase of QCD, the following key
issues need therefore to be addressed: how does the quark condensate 
$\langle \bar{\psi}\psi \rangle$ 
change with temperature and/or baryon density? And are there
systematic changes of hadronic spectral functions in a dense and hot medium,
which would indicate changes of the QCD vacuum structure?

The mass scale given by the gap $\Delta \sim 4 \pi f_{\pi}$ offers a
natural separation between "light" and "heavy" (or, correspondingly, "fast" and
"slow") degrees of freedom. Such a separation of scales is the basis of
effective field theory in which the active light particles are
introduced as collective degrees of freedom, while the
heavy particles are frozen and treated as (almost) static sources. The dynamics
is described by an effective Lagrangian which incorporates all relevant
symmetries of the underlying fundamental theory. This effective Lagrangian is
constructed as follows.

a) The elementary quarks and gluons of QCD are replaced by Goldstone
bosons. They are represented by a matrix field $U (x) \in
SU(2)$ which collects the three isospin components $\pi_a (x)$ of 
the Goldstone pion. A
convenient choice of coordinates is
\begin{equation}
U (x) = \exp[i \tau_a \phi_a(x)]~~ ,
\end{equation}
with $\phi_a = \pi_a/f$ where the pion decay constant $f$ in the chiral limit
provides a suitable normalisation.

b) The QCD Lagrangian (1) is replaced by a chiral effective Lagrangian which
involves the field $U (x)$ and its derivatives:
\begin{equation}
{\cal L}_{QCD} \to {\cal L}_{eff} (U, \partial_{\mu} U, ... ,
\Psi_N, ...) .
\end{equation}
It also involves the coupling of the Goldstone bosons to nucleons treated
as heavy fermion fields $(\Psi_N(x))$. Goldstone bosons interact weakly
at low energy and momentum. The low-energy expansion of ${\cal L}_{eff}$
is therefore organised as an expansion in powers of $\partial_{\mu} U$.

c) Short-distance dynamics which remains unresolved at momentum scales below
the mass gap $\Delta$ is described by approriately adjusted contact
interactions.

The resulting effective Lagrangian has the form
\begin{equation}
{\cal L}_{eff} = {\cal L}_\pi + {\cal L}_{\pi N} + {\cal L}_{NN}~~.
\end{equation}
Goldstone bosons and their non-linear interactions are represented by
${\cal L}_\pi$, its leading term being the non-linear sigma model,
${\cal L}^{(2)} = (f^2/4) 
Tr [ \partial_{\mu} U^{\dagger} \partial^{\mu} U]$, plus a symmetry
breaking mass term. ${\cal L}_{\pi N}$ includes nucleons and their 
interactions with Goldstone pions, with leading vector and axial vector 
(derivative) couplings plus higher order non-linear terms. Complete 
expressions can be found in ref.\cite{15}. The terms ${\cal L}_\pi
+ {\cal L}_{\pi N}$ generate low-energy interactions relevant for 
$\pi\pi$ and $\pi N$ scattering, as well as the long- and intermediate-range 
$NN$ interactions involving one- and two-pion exchange processes.
The short distance $NN$ dynamics, not resolved in detail at momentum
scales small compared to the gap $\Delta$, is encoded in contact
interaction terms denoted by ${\cal L}_{NN}$.

The systematic low-energy expansion of the S-matrix generated by 
${\cal L}_{eff}$ is called chiral perturbation theory. The small 
expansion parameter is $Q/4\pi f_\pi$, where $Q$ stands generically for
the three-momentum or energy of the Goldstone bosons, for the pion mass
$m_\pi$, or for the Fermi momentum $p_F$ of the nuclear system (in which
case one refers to ``in-medium'' chiral perturbation theory).
 
\subsection{Thermodynamics of the chiral condensate}

Let us return to the partition function (18). The Hamiltonian density 
${\cal H}$ of QCD is expressed in terms of the relevant
degrees of freedom in the hadronic phase, derived from the
chiral effective Lagrangian ${\cal L}_{eff}$. The Hamiltonian has a mass
term,
\begin{equation}
\delta{\cal H } = \bar{\psi}m\psi = m_u~\bar{u}u + m_d~\bar{d}d~+~...~~,
\end{equation}
so that ${\cal H} = {\cal H}_0 + \delta{\cal H}$, with ${\cal H}_0$
representing the massless limit.

Now assume a homogeneous medium and consider the pressure (or the free
energy density)
\begin{equation}
P(T,V,\mu) = -{\cal F}(T,V,\mu) = \frac{T}{V}ln{\cal Z}~~.
\end{equation}
The derivative of $P$ with respect to a quark mass $m_q$ of given
flavour $q = u,d$ obviously produces the in-medium quark
condensate, the thermal expectation value $\langle \bar{q}q\rangle_T$
as a function of temperature and chemical potentials. Subtracting
vacuum quantities one finds
\begin{equation}
\langle \bar{q}q\rangle_{T,\rho} = \langle \bar{q}q\rangle_0
- \frac{dP(T,V,\mu)}{dm_q}~~,
\label{con}
\end{equation}
where $\langle \bar{q}q\rangle_0$ refers to the vacuum condensate
taken at $T = 0$ and $\mu = 0$. The $\mu$-dependence of the condensate
is converted into a density dependence via the relation $\rho = \partial
P/\partial\mu$ at fixed $T$.
Using the Gell-Mann, Oakes, Renner relation (11), one can rewrite 
eq.(\ref{con}) as
\begin{equation}
\frac{\langle \bar{q}q\rangle_{T,\rho}}{\langle \bar{q}q\rangle_0} =
1 + \frac{1}{f_{\pi}^2}\frac{dP(T,\mu)}{dm_{\pi}^2}~~.
\label{con2}
\end{equation}
The task is therefore to investigate how the equation of state changes,
at given temperature and baryon chemical potential, when varying the 
quark mass (or, equivalently, the squared pion mass).

For a system of nucleons interacting with pions, the total 
derivative of the pressure with respect to $m_{\pi}^2$ reduces to
\begin{equation}
\frac{dP}{dm_{\pi}^2} = \frac{\partial P}{\partial m_{\pi}^2}
+ \frac{\partial M_N}{\partial m_{\pi}^2}\frac{\partial P}{\partial M_N}
= \frac{\partial P}{\partial m_{\pi}^2} - \frac{\sigma_N}{m_{\pi}^2}
\rho_S(T,\mu)~~,
\label{DP}
\end{equation}
with the scalar density $\rho_S = -\partial P/\partial M_N$ and the 
sigma term $\sigma_N = m_q\partial M_N/\partial m_q = m_{\pi}^2 
\partial M_N/\partial m_{\pi}^2$.

Next, consider the limit of low baryon density $\rho$ at zero temperature,
$T = 0$. In this case nucleons are the only relevant degrees of freedom.
At sufficiently low density, i.e. when the average distance between two 
nucleons exceeds by far the pion Compton wavelength, the nucleon mass remains
at its vacuum value $M_N$, and one can neglect $NN$ interactions. The pressure
is that of a free Fermi gas of nucleons, subject only to the Pauli principle.
Returning to eqs.(27,28), we have
\begin{equation}
\frac{\langle \bar{q}q\rangle_\rho}{\langle \bar{q}q\rangle_0} =
1 - \frac{\sigma_N}{m_\pi^2 f_\pi^2}\rho_S~~,
\end{equation}
with the scalar density
\begin{equation}
\rho_S = -\frac{\partial P}{\partial M_N} =
4\int_{|\vec{p}|\leq p_F} \frac{d^3p}{(2\pi)^3}
\frac{M_N}{\sqrt{\vec{p}\,^2 + M_N^2}}~~,
\end{equation}
for a system with equal number of protons and neutrons, i.e. with degeneracy
factor $d=4$ from spin and isospin. The Fermi momentum $p_F$ is related to
the baryon density by $\rho = 2p_F^3/(3\pi^2)$.
At low baryon densities with $p_F^2 \ll M_N^2$, the difference between $\rho_S$
and $\rho$ can be neglected, so that
\begin{equation}
\frac{\langle \bar{q}q\rangle_\rho}{\langle \bar{q}q\rangle_0} \approx
1 - \frac{\sigma_N}{m_\pi^2 f_\pi^2}\rho~~.
\end{equation}
The consequences implied by this relation are quite remarkable. 
Using the empirical value $\sigma_N\simeq 50$ MeV of the nucleon 
sigma term one observes that the chiral condensate at normal nuclear
matter density, $\rho = \rho_0 = 0.16~fm^{-3}$, is expected to 
decrease in magnitude to less than $2/3$ of its vacuum value, a significant 
effect that should have observable consequences already in the bulk parts
of ordinary nuclei.

In the limit of low temperature and at vanishing baryon 
density $(\rho = 0)$, pions are the only thermally active degrees of
freedom in the partition function (19). This case is well controled
by chiral perturbation theory. The combined low-temperature and low-density
behaviour is summarised as
\begin{equation}
\frac{\langle \bar{q}q\rangle_\rho}{\langle \bar{q}q\rangle_0} \approx
1 - \frac{T^2}{8f_{\pi}^2} - 0.37\left(\frac{\sigma_N}{50 MeV}\right)
\frac{\rho}{\rho_0} + ... ~~,
\end{equation}
showing a rather weak leading dependence of the chiral condensate 
on temperature, whereas its density dependence is more pronounced.

\subsection{A model}

Consider now a schematic model for the hadronic phase of QCD, starting
from the effective Lagrangian (23). The short distance
dynamics is absorbed in $NN$ contact terms,
\begin{equation}
{\cal L}_{NN} = \frac{G_S}{2}(\bar{N}N)^2 - 
\frac{G_V}{2}(\bar{N}\gamma_{\mu} N)^2 + ...~~,
\end{equation}  
with the coupling strength parameters fixed to reproduce ground state
properties of normal nuclear matter. What we have in mind here 
is a variant of relativistic mean field theory combined with 
``soft'' pion fluctuations, generated by ${\cal L_{\pi}} +  {\cal L}_{\pi N}$
and treated in the framework of chiral perturbation theory.

Using two-loop thermal field theory in order to perform a self-consistent
calculation of the pressure $P(T,\mu)$ in this model, one can deduce  
the chiral condensate as a function of temperature and baryon density
following eq.(27). This calculation generates temperature dependent
mean fields for the nucleons at the same time as it treats thermal pion
fluctuations with leading $\pi\pi$ interactions. The pressure equation
takes the form
\begin{equation}
P(T,\mu) = P_N(T,\mu^*,M_N^*) + P_\pi(T,\mu^*,M_N^*)
-\frac{G_S}{2}\rho_S^2 + 
\frac{G_V}{2}\rho^2~~~,
\end{equation}
with nucleon and pion contributions $P_{N,\pi}$ respectively depending on
the effective nucleon mass $M_N^* = M_N - G_S~\rho_S$
and the shifted baryon chemical potential $\mu^* = \mu - G_V\rho$.
The baryon and scalar densities are determined as
\begin{equation}
\rho = \frac{\partial P}{\partial\mu} =
\frac{\partial(P_N + P_\pi)}{\partial\mu^*}, ~~~~
\rho_S = \frac{\partial P}{\partial M_N} =
\frac{\partial(P_N + P_\pi)}{\partial M_N^*}~~~.  
\end{equation}
Fixing $G_{S,V}$ to the energy per particle and the equilibrium 
density of cold nuclear matter, one can map out the nuclear equation
of state first at low temperature and density. Reproducing known physics
in this domain is a prerequisite for extrapolating into more extreme
regions.

\begin{figure}[t]
\begin{center}
\parbox{10cm}{
\includegraphics[width=10cm,clip=true]{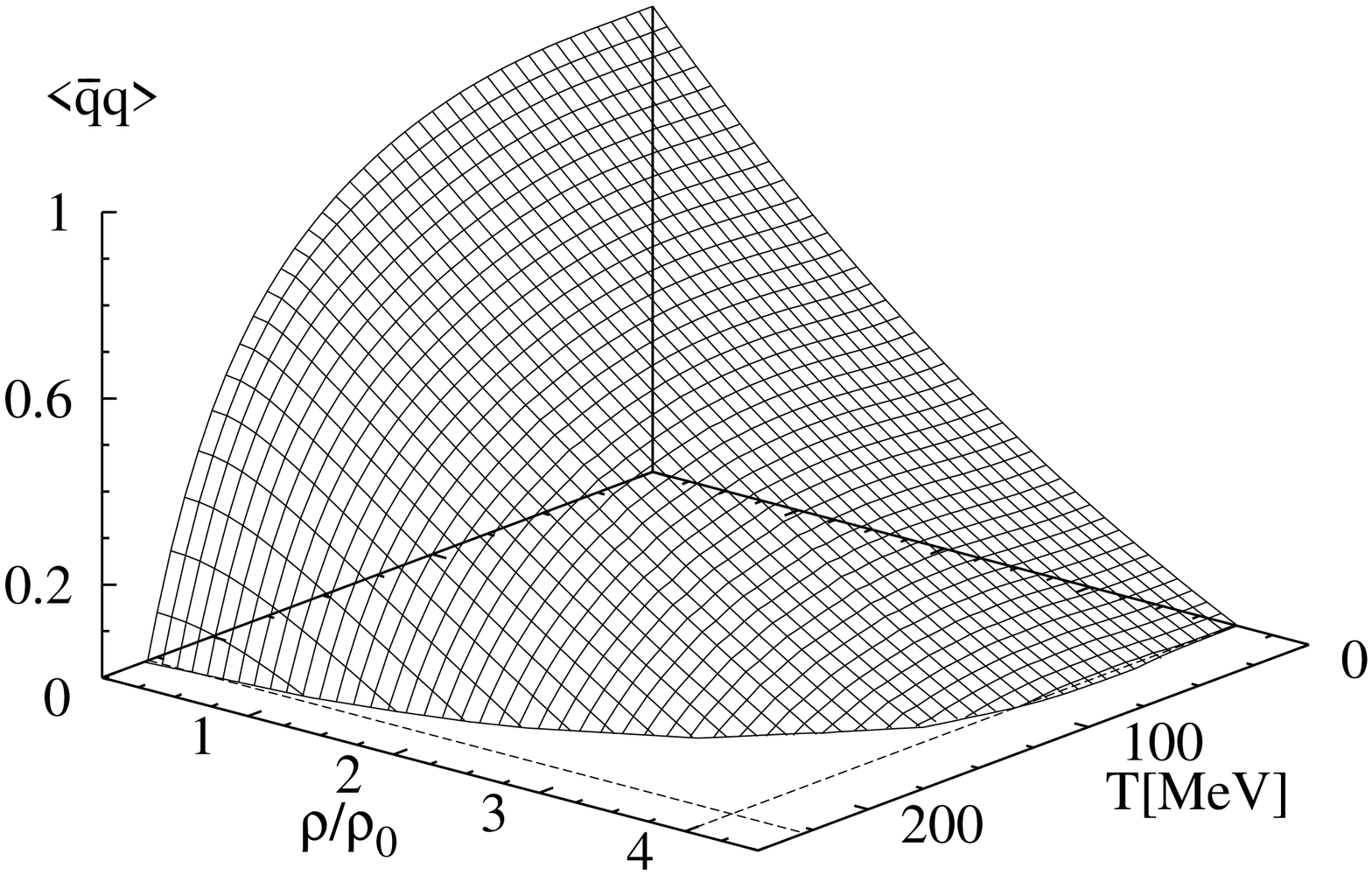}}
\end{center}
\caption{Chiral condensate (in units of its vacuum value) as a function
of temperature and baryon density ($\rho_0 = 0.16 fm^{-3}$ is the density
of normal nuclear matter).}
\label{fig:qqT}
\end{figure}

We now return to the evaluation of the chiral condensate. The dependence
of $P(T,\mu)$ on the pion mass is explicit in the thermal pion Green
function and implicit through the nucleon mass. The result \cite{16} 
for $\langle \bar{q}q\rangle_{T,\rho}$ is shown 
in Fig.\ref{fig:qqT}. The temperature dependence is reminiscent of 
the corresponding lattice QCD result. 

At low baryon density the linear behaviour of eq.(31) is recovered. 
Pionic fluctuations, calculated up to three-loop order with inclusion
of two-pion exchange effects, help maintain this approximately
linear dependence not only up to $\rho\simeq\rho_0$, but even slightly beyond.
Up to this point we can conclude that the magnitude of the quark
condensate at normal nuclear matter density is expected to be reduced by 
about one third from its vacuum value, whereas the temperature dependence
is far less pronounced, at least up to $T\le 100 MeV$.

The GOR relation (11) continues to hold \cite{17} in matter at finite 
temperature $T < T_c$ and at finite density, when reduced to a statement
about the $\it{time}$ component, $A_a^0 = \psi^{\dagger}\gamma_5 (\tau_a/2)
\psi$, of the axial current (5). Introducing the in-medium pion 
decay constant, $f_{\pi}^*(T,\rho)$, through the in-medium analogue 
of eq.(10), one finds  
\begin{equation}
f_{\pi}^*(T,\rho)^2~m_{\pi}^*(T,\rho)^2 = -(m_u + m_d)
\langle\bar{q}q\rangle_{T,\rho} + ... ~~
\label{inmedGOR}
\end{equation}
to leading order in the quark mass. It turns out that the in-medium pion mass
$m_{\pi}^*$ (actually the average of the $\pi^+$ and $\pi^-$
masses) is protected by the pion's Goldstone boson nature and 
not much affected 
by the thermal environment. The ``melting'' of the condensate by heat or
compression translates primarily into the in-medium change of the pion decay 
constant.

\subsection{Hadronic mass spectra in dense and hot matter}

The fact that the ``chiral gap'' $\Delta \sim 4\pi f_{\pi}^*(T,\rho)$ 
decreases when
thermodynamic conditions change towards chiral restoration, should imply
observable changes in the spectral distributions of hadrons in 
dense and hot matter. A great amount of theoretical activities and 
experimental searches is focused on this theme. The present situation can
be briefly summarised as follows \cite{18}:

a) The spectral distribution of scalar-isoscalar $(\pi\pi)$ excitations 
is expected to show a strong enhancement as critical conditions are
approached \cite{HK}. Whether such an enhancement can actually be observed
in dedicated experiments is an important issue of presently ongoing 
discussions.

b) Spectral functions of vector and axial vector excitations (the $\rho$
and $a_1$ mesons) tend to become degenerate as $T$ approaches $T_c$ \cite{19}.
At finite density the spectrum of $\rho$ mesons shows substantial broadening
as a consequence of collisions with baryons and mesons in the medium, however,
so that the connection with the chiral condensate gets largely diluted
\cite{18,20}.
The re-distribution of strength in vector channels, primarily by the 
broadening of the $\rho$ meson spectrum, is consistent with the
observed dilepton yields from $Pb + Au$ collisions at CERN \cite{18,21}.

c) The low-energy (s-wave) dynamics of pions in a nuclear system should 
be sensitive to possible in-medium changes of the pion decay constant. 
This is the question that we are now going to address in more detail.

\section{Goldstone bosons in matter}

\subsection{Pion self-energy}

Consider homogeneous nuclear matter at zero temperature with proton density
$\rho_p$ and neutron density $\rho_n$. A pion wave in matter has its
energy $\omega$ and momentum $\vec{q}$ related by the dispersion
equation  
\begin{equation}
\omega^2 - \vec{q}~^2 - m_{\pi}^2 - 
\Pi(\omega,\vec{q}~;\rho_p,\rho_n) = 0~~~.
\end{equation}
The quantity $\Pi$ which summarises all interactions of the pion
with the medium is called the pion polarisation function, or self-energy.
At low densities,
\begin{equation}\label{selfen}
\Pi(\omega,\vec{q}~;\rho_p,\rho_n) = -T_{\pi p}(\omega,\vec{q}~)~\rho_p 
- T_{\pi n}(\omega,\vec{q}~)~\rho_n + ...~~~,
\end{equation}
where $T_{\pi N}(\omega,\vec{q}~)$ are the pion-nucleon forward 
amplitudes taken at the respective energy and momentum. It is convenient 
to rewrite eq.(\ref{selfen}) as
\begin{equation}
\Pi^{(\pm)}(\omega,\vec{q}~;\rho_p,\rho_n) = -T^{+}(\omega,\vec{q}~)~\rho 
\pm T^{-}(\omega,\vec{q}~)~\delta\rho~~~,
\end{equation}
in terms of the isospin-even and isospin-odd pion-nucleon amplitudes, with
$\rho = \rho_p + \rho_n$ and $\delta\rho = \rho_p - \rho_n$,
where we have now specified the self-energies $\Pi^{(\pm)}$ for a $\pi^+$
or $\pi^-$, respectively (for a $\pi^0$ the isospin-odd term drops out).

When inserted into the wave equation (37), the polarisation function (39)
determines the spectrum $\omega(\vec{q}~)$ of pionic modes of excitation
in an infinite nuclear medium. Applications to finite systems, in particular
for low-energy pion-nucleus interactions close to threshold relevant for 
pionic atoms, commonly 
use an energy-independent effective potential (the optical potential).
Such an equivalent potential can be constructed as follows 
\cite{22}. Expand the 
polarisation function for $\omega - m_\pi \ll m_\pi$ and $|\vec{q}~|^2 \ll
m_\pi^2$ around the physical threshold, $\omega = m_\pi$ and $|\vec{q}~| = 0$:
\begin{equation}
\Pi^{(\pm)}(\omega,\vec{q}\,) = \Pi^{(\pm)}(m_\pi,0)
+ \frac{\partial\Pi}{\partial\omega}(\omega-m_\pi) +
 \frac{\partial\Pi}{\partial\vec{q}\,^2}\vec{q}\,^2 ~~~,
\end{equation} 
where the derivatives are taken at threshold. Substituting this expansion
in eq.(37) one obtains
\begin{equation}
\omega^2 - m_\pi^2 - \frac{1}{1-\alpha(\omega)
\frac{\partial\Pi}{\partial\omega^2}}\left[\Pi(m_\pi,0) + \vec{q}\,^2
\left(1+\frac{\partial\Pi}{\partial\vec{q}\,^2}\right)\right] = 0
\end{equation} 
with $\alpha(\omega)=2\omega/(\omega+m_\pi)\simeq 1$ at $\omega\simeq m_\pi$.
By comparison with the Klein-Gordon equation for the pion wave function
$\phi(\vec{r}~)$ in coordinate space,
\begin{eqnarray}\label{kge1}
\left[ \omega^2-m_\pi^2+\vec{\nabla}^2-
2m_\pi U(\vec{r}\,)\right]\phi(\vec r\, )=0\, ~~~,
\end{eqnarray}
the (energy-independent) optical potential $U(\vec{r}~)$ is identified
as 
\begin{equation}
2m_\pi U(\vec{r}\,) = \left(1-\frac{\partial\Pi}{\partial\omega^2}\right)^{-1}
\left[\Pi(m_\pi,0) +\vec{\nabla}\left(\frac{\partial\Pi}{\partial\omega^2}
+\frac{\partial\Pi}{\partial\vec{q}\,^2}\right)\vec{\nabla}\right]~~~,
\label{eq:pot}
\end{equation} 
with all derivatives taken at the threshold point. The wave 
function renormalisation factor $(1-\partial\Pi/\partial\omega^2)^{-1}$ 
encodes the energy dependence of the polarisation function
$\Pi(\omega,\vec{q}\,)$ in the equivalent energy-independent potential 
(\ref{eq:pot}). This potential is expressed in terms of local density
distributions $\rho_{p,n}(\vec{r}~)$ for protons and neutrons, and the
standard prescription $\vec{q}\,^2f(\rho) \rightarrow 
\vec{\nabla}f(\rho(\vec{r}\,))\vec{\nabla}$ is used for the $\vec{q}\,^2$-
dependent parts. In practical calculations of pionic atoms, the Coulomb
potential $V_c$ is introduced by replacing $\omega \rightarrow \omega -
V_c(\vec{r}\,)$, and corrections of higher order beyond the leading terms
(39), resulting from double scattering and absorption, are added.  
 
\subsection{Deeply bound states of pionic atoms}

Recent accurate data on 1s states of a negatively charged pion bound 
to Pb and Sn isotopes~\cite{23} have set new standards and 
constraints for the detailed analysis of s-wave pion interactions with
nuclei. Such deeply bound pionic states owe their existence, with relatively
long lifetimes, to a subtle balance between the attractive Coulomb force
and the repulsive strong $\pi^-$-nucleus interaction in the bulk of
the nucleus. As a consequence, the 1s wave function
of the bound pion is pushed toward the edge of the nuclear surface. Its
overlap with the nuclear density distribution is small, so that the standard
$\pi^- pn \rightarrow nn$ absorption mechanism is strongly suppressed.

The topic of low-energy, s-wave pion-nucleus interactions has a long history
\cite{29,22}, culminating in numerous attempts
to understand a persisting ``missing repulsion'': the standard ansatz 
for the (energy independent) s-wave
pion-nucleus optical potential, given in terms of the empirical $\pi N$
threshold amplitudes times the proton and neutron densities, $\rho_{p,n}$ 
and supplemented 
by important double scattering corrections, still misses the empirically
observed
repulsive interaction by a large amount. This problem has traditionally been
circumvented on purely phenomenological grounds by simply introducing an
unusually large repulsive real part (Re$B_0$) in the $\rho^2$ terms of the 
pion-nucleus optical potential. The arbitrariness of this procedure is
of course unsatisfactory.

This issue has recently been re-investigated~\cite{24} from the point of view
of the specific energy dependence of the pion-nuclear polarisation
operator in a calculation based on systematic in-medium chiral perturbation
theory~\cite{25}. Ref.\cite{24} has also clarified the relationship
to a working hypothesis launched previously~\cite{26,27}: namely that the
extra repulsion needed in the s-wave pion-nucleus optical potential at
least partially reflects the tendency toward chiral symmetry restoration
in dense matter, effectively through a reduction of the in-medium pion decay 
constant, $f_{\pi}\rightarrow f_{\pi}^*(\rho)$. This prescription has
proved remarkably successful in reproducing the systematics over the 
complete pionic atoms data base, using optical potential 
phenomenology~\cite{28}. 

Consider a negatively charged pion interacting with nuclear matter. 
In the limit of very low proton and neutron densities, 
$\rho_p$ and $\rho_n$, the $\pi^-$ self-energy reduces to  
$\Pi(\omega,\vec{q}~;\rho_p,\rho_n) = -[T^{+}(\omega,\vec{q}~)~\rho 
+ T^{-}(\omega,\vec{q}~)~\delta\rho]$,
as in eq.(39). In the long-wavelength limit ($\vec{q}\rightarrow 0$), 
chiral symmetry
(the Tomozawa-Weinberg theorem) implies $T^-(\omega) =
\omega/(2 f_{\pi}^2) + {\cal O}(\omega^3)$. Together with the observed
approximate vanishing of the isospin-even threshold amplitude 
$T^+(\omega = m_{\pi})$, it is clear that $1s$ states of pions bound to
heavy, neutron rich nuclei are a sensitive source of information
for in-medium chiral dynamics.

At the same time, it has
long been known that terms of non-leading order in density (double
scattering (Pauli) corrections of order $\rho^{4/3}$, absorption
effects of order $\rho^2$ etc.) are important \cite{29}. The aim must,
therefore, be to arrive at a consistent expansion of the pion
self-energy in powers of the Fermi momentum $p_{\rm F}$ together
with the chiral low-energy expansion in  $\omega, |\vec{q}\, |$
and $m_\pi$\,.  In-medium chiral effective field theory provides a
framework for this approach. It has been systematically applied
in refs.\cite{24,25}, and its formally more rigorous implementation for 
finite nuclear systems is under way \cite{30}. Double scattering
corrections are fully incorporated. Absorption
effects and corresponding dispersive corrections appear at the
three-loop level and through  short-distance dynamics
parametrised by $\pi N N$ contact terms, not explicitly calculable
within the effective low-energy theory. The imaginary parts
associated with these terms are well constrained  by the
systematics  of observed  widths of pionic atom levels throughout
the periodic table. (We use ${\rm Im} B_0=-0.063 m_\pi^4$ in the
s-wave absorption term, 
$\Delta \Pi^{\rm abs}_{\rm S}=-8\, \pi\,
(1+m_\pi/2 M)\, B_0\, \rho_p\, (\rho_p+\rho_n)$,
and the canonical parameterization of p-wave parts). 
The real part of $B_0$ is still a major
source of theoretical uncertainty. In practice, a good strategy is to
start from ${\rm Re} B_0=0$ (as suggested also by the detailed
analysis of the pion-deuteron scattering length) and then discuss
the possible error band  induced by varying ${\rm B_0}$ within
reasonable limits.

We proceed by using the local density approximation (with gradient
expansion for p-wave interactions) and solve the
Klein-Gordon equation
\begin{eqnarray}\label{kge}
\Big[ \Big(\omega-V_c(\vec r\,)\Big)^2+\vec\nabla^2-m_\pi^2-
\Pi\Big(\omega-V_c(\vec r\,);
\rho_p(\vec r\, ),\rho_n(\vec r\,)\Big)\Big]\phi(\vec r\, )=0\,.
\end{eqnarray}
Note that the explicit energy dependence of $\Pi$ requires that
the Coulomb potential $V_c(\vec r\, )$ must be introduced in the
canonical gauge invariant way wherever the pion energy $\omega$
appears. This  is an important feature that has generally been
disregarded in previous analysis. The connection with the 
energy-independent optical potential (\ref{eq:pot}) and with possible 
in-medium renormalisation effects will be discussed later in this section. 

\begin{figure}[t]
\begin{center}
\parbox{7.0cm}{
\includegraphics[width=7.0cm,clip=true]{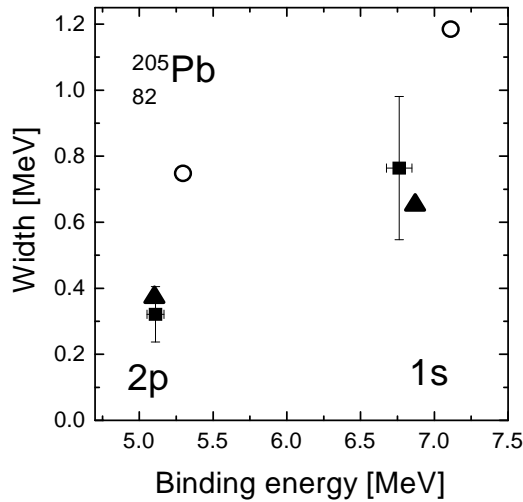}}
\end{center}
\caption{Binding energies and widths of pionic $1s$ and $2p$ states in
  $^{205}$Pb. Experimental data from~\cite{23}. Full  triangles:
  results of two-loop in-medium chiral perturbation theory, keeping
  the explicit energy dependence in the s-wave polarization
  operator. Open circles: energy independent potential as described in
  text (see ref.~\cite{24} for details). Note that ${\rm Re}B_0=0$ in
  both cases.}
\label{fig:pb}
\end{figure}

With input specified in detail in ref.\cite{24}, we have solved
eq.~(\ref{kge}) with the explicitly energy dependent pion
self-energy, obtained in two-loop in-medium chiral perturbation
theory for the s-wave part, adding the time-honored
phenomenological p-wave piece. The results for the binding energies
and widths of $1s$ and $2p$ states in pionic $^{205}$Pb are shown
in Fig.~\ref{fig:pb} (triangles). Also shown for comparison is the
outcome of a calculation using a "standard" phenomenological
(energy independent) s-wave optical potential,
\begin{eqnarray}\label{pis}
\Pi_{\rm S}=-T_{\rm eff}^+\, \rho-T_0^-\,\delta
\rho+\Delta\Pi_{\rm S}^{\rm abs}\,,
\end{eqnarray}
with  $T_{\rm eff}^+=T_0^+-\frac{3 p_{\rm F}(\vec r)}{8\, \pi^2}
\, [(T_0^+)^2+2\, (T_0^-)^2]$ including double scattering effects
and the amplitudes
$T_0^{\pm}\equiv T^{\pm}(\omega=m_\pi)$ taken fixed at their
threshold values. This approach fails and shows the "missing
repulsion" syndrome, leading  to a substantial overestimate of the
widths. Evidently, a mechanism is needed to reduce the overlap of
the bound pion wave functions with the nuclear  density
distributions. The explicit energy  dependence in $T^{\pm}$
provides such a mechanism: the replacement
$\omega \to \omega-V_c(\vec r\,)>m_\pi$ increases the repulsion in $T^-$
and  disbalances the "accidental" cancellation between the $\pi N$
sigma term $\sigma_{ N}$ and the range term proportional to
$\omega^2$ in $T^+$ (see eq.(46)), such that $T^+(\omega-V_c)<0$ (repulsive).
Uncertainties in ${\rm Re} B_0$, in the radius and shape of the
neutron density distribution, and in the input for the sigma term
$\sigma_N$ have been analysed in ref.~\cite{24}. Their combined
effect falls within the experimental errors in Fig.~\ref{fig:pb}.

\begin{figure}
\begin{center}
\parbox{7.0cm}{
\includegraphics[width=7.0cm,clip=true]{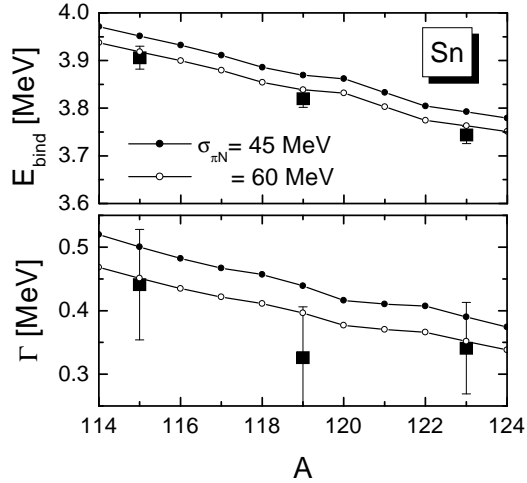}}
\end{center}
\caption{
Binding energies and widths of pionic $1s$ states in Sn isotopes. The
curves show predictions~\cite{31} based on the explicitly  energy
dependent pionic s-wave polarization operator calculated in two-loop
in-medium chiral perturbation theory~\cite{24,25}. The sensitivity to
the $\pi N$ sigma term (input) is also shown. Data from ref.~\cite{23}.
}
\label{fig:sn}
\end{figure}

Using the same (explicitly energy dependent) scheme we have
predicted binding energies and widths for pionic $1s$ states bound
to Sn isotopes. These calculations
include  a careful assessment of uncertainties in neutron
distributions for those isotopes. Results are shown in Fig.~\ref{fig:sn} in
comparison with experimental data (which were actually
reported after the calculations). This figure also gives an
impression of the sensitivity with respect to variations of the
(input) $\pi N$ sigma term.

\subsection{Concluding remarks: fingerprints of chiral restoration ?}

We finally come to an important question of interpretation: do we
actually "observe" fingerprints of (partial) chiral symmetry
restoration in the high-precision data of deeply bound pionic
atoms with heavy nuclei, as anticipated in refs.\cite{26,27}? Is
this observation related to the "missing s-wave repulsion" that
has been recognized (but not resolved in a consistent  way) by
scanning the large amount of already existing pionic atom data?

To approach this question, recall that pionic atom calculations
are traditionally done with \emph{energy-independent}
phenomenological optical potentials instead of explicitly energy
dependent pionic polarisation functions. Let us examine the
connection between these two seemingly different approaches by
illustrating the leading-order driving mechanisms.

Consider a zero momentum $\pi^-$ in low density 
matter. Its in-medium dispersion equation at $\vec {q}=0$ is
$\omega^2-m_\pi^2-\Pi(\omega)=0$\,. The chiral low-energy
expansion of the off-shell amplitudes $T^{\pm}(\omega)$ at $\vec {q}=0$
implies leading terms of the form:
\begin{eqnarray}
\label{t}
T^+(\omega)=\frac{\sigma_N-\beta\, \omega^2}{f_\pi^2}\,,
\quad T^-(\omega)=\frac{\omega}{2\, f_\pi^2}\,,
\end{eqnarray}
with the pion-nucleon sigma term $\sigma_N \simeq 50$ MeV and 
$\beta\simeq \sigma_N/m_\pi^2$.
Expanding $\Pi(\omega)$ around the threshold, $\omega=m_\pi$\,,
and returning to eq.(43)
we identify the commonly used effective (energy-independent)
s-wave optical potential $U_{\rm S}$ as:
\begin{eqnarray}\label{us}
2\, m_\pi\, U_{\rm S}=\frac{\Pi(\omega=m_\pi,\vec
q=0)}{1-\partial\Pi/\partial\omega^2}\,,
\end{eqnarray}
where $\partial\Pi/\partial\omega^2$ is taken at $\omega=m_\pi$\,.
Inserting $\Pi(\omega,\vec{q}=0~;\rho_p,\rho_n) = -T^{+}(\omega)~\rho 
- T^{-}(\omega)~\delta\rho$, and assuming $\delta\rho\ll \rho$ one finds:
\begin{eqnarray}
\label{us2}
U_{\rm S}\simeq -\frac{\delta \rho}{4\, f_\pi^2}\,
\left(1-\frac{\sigma_N\, \rho}{m_\pi^2\, f_\pi^2}\right)^{-1} =
-\frac{\delta \rho}{4\, f_\pi^{* 2}(\rho)}\,,
\end{eqnarray}
with the replacement $f_\pi\to f_\pi^*(\rho)$ of the pion decay
constant representing the in-medium wave function renormalization.
The expression (\ref{us2}) is just the one proposed previously in
ref.\cite{26} on the basis of the relationship between the
in-medium changes of the chiral condensate 
$\langle\overline{q}\, q\rangle$
and of the pion decay constant associated with the time
component of the axial current (see eq.(36)). The explicitly energy dependent
chiral dynamics represented by $\Pi(\omega)$ "knows" about these
renormalization effects. Their  translation into an equivalent,
energy-independent potential implies  $f_\pi\to f_\pi^*(\rho)$ as
given in eq.~(\ref{us2}). This statement holds to leading order.
It still needs to be further explored in detail whether this 
interpretation is stable against higher order corrections.

The appearance of $f_{\pi}^*(\rho)$ in the denominator of the
equivalent energy-independent potential $U_S$ is what we refer
to as ``fingerprints of chiral restoration'', in the sense that
$f_{\pi}$ is an order parameter of spontaneous chiral symmetry
breaking in QCD and it tends to decrease with increasing baryon density.

\vspace{1cm}

The author gratefully acknowledges the kind hospitality extended to
him by Atsushi Hosaka, Teiji Kunihiro and Hiroshi Toki during his
visits to Osaka and Kyoto.

\end{document}